# Magnetization-control and transfer of spin-polarized Cooper pairs into a half-metal manganite

A. Srivastava[1], L. A. B. Olde Olthof[1,2], A. Di Bernardo[1], S. Komori[1], M. Amado[1], C. Palomares-Garcia[1], M. Alidoust[3], K. Halterman[4], M. G. Blamire[1], and J. W. A. Robinson[1]*

[1]*Department of Materials Science and Metallurgy, University of Cambridge, 27 Charles Babbage Road, Cambridge CB3 0FS, United Kingdom*

[2]*Faculty of Science and Technology and MESA+ Institute for Nanotechnology, University of Twente, 7500 AE Enschede, The Netherlands*

[3]*Department of Physics, K.N. Toosi University of Technology, Tehran 15875-4416, Iran*

[4]*Michelson Lab, Physics Division, Naval Air Warfare Center, China Lake, California 93555, USA*

The pairing state and critical temperature ($T_C$) of a thin s-wave superconductor (*S*) on two or more ferromagnets (*F*) are controllable through the magnetization-alignment of the *F* layers. Magnetization misalignment can lead to spin-polarized triplet pair creation, and since such triplets are compatible with spin-polarized materials they are able to pass deeply into the *F* layers and so, cause a decrease in $T_C$. Various experiments on $S/F_1/F_2$ "triplet spin-valves" have been performed with the most pronounced suppression of $T_C$ reported in devices containing the half-metal ferromagnet (HMF) $CrO_2$ ($F_2$) albeit using out-of-plane magnetic fields to tune magnetic non-collinearity [Singh *et al.*, Phys. Rev. X 5, 021019 (2015)]. Routine transfer of spin-polarized triplets to HMFs is a major goal for superconducting spintronics so as to maximize triplet-state spin-polarization. However, $CrO_2$ is chemically unstable and out-of-plane fields are undesirable for superconductivity. Here, we demonstrate low field (3.3 mT) magnetization-tuneable pair conversion and transfer of spin-polarized triplet pairs to the chemically stable mixed valence manganite $La_{2/3}Ca_{1/3}MnO_3$ in a pseudo spin-valve device using in-plane magnetic fields. The results match microscopic theory and offer full control over the pairing state.

PACS numbers: 74.78.Na, 74.20.-z, 74.25.Ha

## I. INTRODUCTION

Superconducting spintronics represents a new paradigm for information processing involving the coexistence of spin-polarization and superconducting phase coherence [1–3]. Conventional s-wave superconductivity involves the condensation of spin-singlet electron pairs with antiparallel spins. Although singlet pairs are energetically unstable in a ferromagnet, they are able to penetrate a transition metal ferromagnet (*F*) at a superconductor/ferromagnet (*S/F*) interface over distances of a few nanometers [4–10], but without transferring a net spin. Furthermore, singlet pairs are blocked at a *S* interface with a half metallic ferromagnet (HMF) as there are no available states for one of the two spins of a pair to enter since the Fermi energy for the minority-spin electrons falls within a gap.

Electrons pairs in the *p*-wave superconducting compound $Sr_2RuO_4$ [11] have parallel spins and so such spin-triplet pairs carry a net spin in addition to charge. However, the extreme sensitivity of *p*-wave superconductivity to structural and electronic disorder, creates major obstacles to the development of *p*-wave devices [12]. Spin-triplet pairs with parallel spins, but *s*-wave symmetry may form at magnetically inhomogeneous s-wave *S/F* interfaces [1–3]. Since such pairs are compatible with fully spin-polarized materials, their routine creation and transfer to HMFs would open up exciting opportunities for applications in superconducting spintronics where high spin-polarization and long spin-flip scatter lengths are desirable.

---

* jjr33@cam.ac.uk

Spin-polarized triplet pairs form via spin mixing and spin-rotation processes at $S/F$ interfaces [13]. At homogeneously magnetized $S/F$ interfaces or within magnetically collinear $S/F_1/F_2$ spin-valves, spin-singlet pairs experience a spatially constant exchange field that acts differentially on the antiparallel spins of a pair, causing transformation to a spin-zero triplet state (spin-mixed state). A rotation of the magnetization at a $S/F$ interface or within a $S/F_1/F_2$ spin-valve has the effect of transforming spin-zero triplets to pairs with a parallel projection of spin (spin-rotation). For $S/F_1/F_2$ spin-valves where $S$ and $F_1$ ("spin-mixer" layer) are thinner than the spin-singlet coherence length (40 nm in Nb [14] and 1 nm in Co, Fe and Ni [15, 16]), spin-polarized triplet pair creation leads to an effective leakage of superconductivity from $S$ into $F_2$ and a reduction of the critical temperature ($T_C$). "Triplet spin-valves" (TSVs) are therefore sensitive devices for investigating singlet-to-triplet pair conversion [17–20].

Experiments over the past few years have mainly focused on magnetization-control of triplet pair creation in $S/F/S$ Josephson devices and TSVs. In $S/F/S$ devices various symmetric spin-mixer layers have been added to the $S/F$ interfaces, including rare earth magnetic spirals [21, 22], antiferromagnets [23], Heusler alloys [24], and transition metal ferromagnets [25–29]. Similarly in $S/F_1/F_2$ TSVs, $F_{1,2}$ metals [30–33] or $F$ metals ($F_1$) in combination with the HMF $CrO_2$ ($F_2$) [34] have been successfully demonstrated. See also related works on $F/S/F$ spin-valves [35–37] and spectroscopy experiments on various S/F systems experiments [38–48].

The most pronounced suppressions of $T_C$ was reported in a MoGe/Ni/Cu/$CrO_2$ TSV in which out-of-plane magnetic fields created a misalignment between the magnetizations of Ni and $CrO_2$ [34]; the largest suppression of $T_C$ was close to -800 mK with a constant out-of-plane magnetic field of 2 T. This pioneering work extended previous experiments that demonstrated Josephson coupling across $CrO_2$ [49] (see also [27,35]) in devices that did not contain intentional spin-mixer layers at the $S$/HMF interfaces. However, $CrO_2$ is chemically unstable and so there is a need to identify alternative HMFs in which thin films can be grown and combined with various S/F structures with enhanced chemical stability.

Mixed valance manganites ($La_{1-x}Ae_xMnO_3$, where Ae is an alkaline earth) such as $La_{1-x}Sr_xMnO_3$ (LSMO) and $La_{2/3}Ca_{1/3}MnO_3$ (LCMO) are highly attractive alternatives to $CrO_2$ since they are chemically stable and their relatively narrow spin up and spin down conduction bands are completely separated leading to HMF behaviour at low temperatures[50, 51]. In this *Article*, we report TSV with Nb/Cu/Py/Au/LCMO layers in which a non-monotonic dependence of $T_C$ on the relative magnetization angle (θ) between Py(NiFe) and LCMO is observed, thus demonstrating pair conversion and transfer of spin-polarized triplets to LCMO. Recently, we detected Josephson coupling across thin (< 30 nm) layers of LCMO [52], but without intentional spin-mixers at the $S$/LCMO interfaces. Related experiments that probe spectroscopic signatures triplet pairing in $S$/LCMO structures have also been reported [42–44, 53], but again without intentional spin-mixer layers. The motivation of the work reported here was to investigate magnetization-control of triplet pair creation and transfer to LCMO, which is fundamental to the development triplet superconductivity based on mixed valance manganites. Furthermore, we wanted to demonstrate triplet pair creation in TSVs with small in-plane magnetic fields to avoid complications due voritices that will be present in TSV that require large out-of-plane magnetic fields

## II. EXPERIMENT

We prepared Nb(25nm)/Cu(5nm)/Py(3.5nm)/Au(5nm)/LCMO(120nm) TSVs in several stages. Epitaxial (002) LCMO was grown from a stoichiometric target by pulse laser deposition (PLD) (KrF laser, wavelength λ = 248 nm) on 5 mm x 5mm single crystal $SrTiO_3$ (001) at a growth temperature of 800 °C in flowing $N_2O$ at 130 mTorr with a pulse fluence of 1.5 J/cm$^2$ for 15 minutes and repetition rate of 2 Hz, then 30 minutes at 3 Hz. The films were annealed *in situ* at the same temperature in oxygen (46 kPa) for 8 hours and cooled to room temperature at a rate of 10 °C/min. High resolution X-ray diffraction (Fig. 1S) confirmed single (002) orientation of LCMO with rocking curves on the (002), (004), (006) and (008) Bragg peaks showing full width at half maximum values of 0.12°, 0.18°, 0.209° and 0.227°, respectively. The c-axis lattice parameter was determined to be 7.670±0.002 Å, consistent with powder diffraction simulations [54]. Au was deposited on LCMO at room temperature using a fluence of 2.5 J/cm$^2$ for 3 minutes at 5 Hz in 30 mTorr of Ar (Au was chosen due to its oxidation resistance and limited solubility with Ni). Au/LCMO bilayers were then transferred in air to an ultrahigh vacuum sputtering system with a base pressure of 3 x 10$^{-9}$ mBar and Nb/Cu/Py trilayers were deposited on Au/LCMO in Ar at 1.5 Pa while rotating below stationary magnetrons. The surface of Au was cleaned in situ by Ar ion plasma etching (-0.6 kV extraction energy and 1 kV ion energy) and different etching times in the 0-5 minute range were investigated. During the sputter process, samples experienced a constant in plane magnetic field of approximately 50 mT.

Control samples of Au(5nm)/LCMO(120nm) and Nb(25nm)/Cu(5nm)/Py(3.5nm)/Au(5nm) were prepared on 5 mm x 5 mm area STO (001) and single crystal silicon substrates, respectively, to characterize the isolated magnetic properties of LCMO and Py. Magnetization $M$ versus applied field $H$ is shown in Fig. 1(a,b) at 10 K. The $M(H)$ of LCMO shows an easy-plane behaviour with an in-plane saturation field ($H_S$) of 50 mT and coercivity ($H_C$) of 20 mT. In the Supplemental Materials [55] we also show that (Fig. 4S) the LCMO is magnetically isotropic in-plane at 10 K. In comparison, the Py shows some in plane anisotropy with an easy axis (EA, defined as 90°) parallel to the field direction during growth and $H_C$ of 1.8 mT and a harder axis (HA, defined as 0°) at a right angle to the EA with $H_C$ = 1.1 mT. The volume saturation magnetizations of LCMO and Py were 470±15 emu/cm$^3$ and 650±25 emu/cm$^3$ respectively, which are similar to values reported elsewhere [For LCMO see [56] and [57] for Py].

Figure 1(c) shows $M(H)$ of the TSV at 10 K where $M$ is dominated by the 120-nm-thick LCMO layer and so, for comparison easy-axis $M(H)$ loop is plotted for the Nb/Cu/Py/Au control (reproduced from Fig. 1(b)). The $M(H)$ loops show that the TSV magnetization state is parallel (P) beyond ±30 mT, and a reversal field of -1.8 mT switches the Py moment to achieve an antiparallel (AP) state.

Resistance vs temperature $R(T)$ measurements of the TSVs were performed using a four-point current-bias technique on unpatterned samples in a pulse-tube measurement system. The $T_C$ was defined as the temperature corresponding to 50% of the normal state resistance. We note that care was taken to ensure that the bias-current (10 µA) had no effect on $R(T)$ through the superconducting transition and therefore that the $T_C$ was current-bias independent (meaning the bias-current is not large enough for vortex-induced voltages to dominate the transport signal). In all cases, $R(T)$ did not show anomalies (e.g. steps) through the superconducting transition.

The effect of in-plane magnetization configuration on $T_C$ was investigated by measuring $R(T)$ though the superconducting transition as a function of the relative magnetization angle (θ) between LCMO and Py. The $T_C(θ)$ measurement routine is illustrated in Fig. 2(a) and described here: (1) at 10 K an external field of 100 mT was applied along the HA of Py to magnetize LCMO and Py (along 0°); (2) a magnetic field of -3.3 mT (<$H_C$ of LCMO) was then applied along the HA of Py to reverse the Py moment and obtain the AP-state (along 180°) and from $R(T)$ in cooling and warming $T_C(180°)$ was obtained; (3) the sample warmed to 10 K and rotated in-plane to an angle θ in a constant field of amplitude 3.3 mT and from $R(T)$ in cooling and warming $T_C(θ)$ was obtained. Stage (3) was repeated at 20° increments to obtain $T_C(θ)$ between 0° - 180°. We note that a field of -3.3 mT was large enough to fully magnetize Py in all in-plane field directions without altering the remnant state of LCMO.

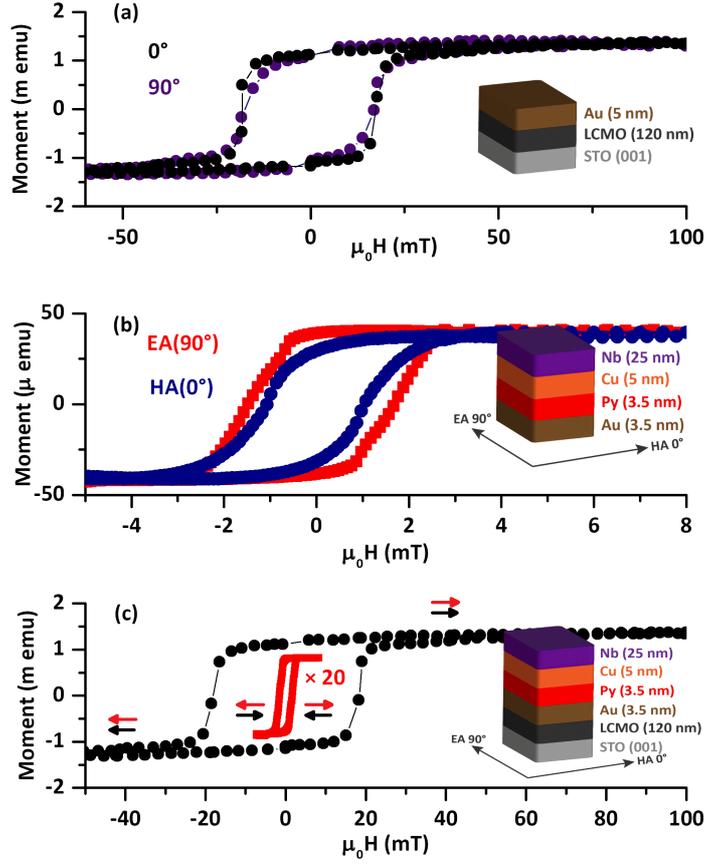

**Fig. 1.** (a) *M*(*H*) loops of LCMO for orthogonal in-plane fields at 10 K. (b) *M*(*H*) of Py with the field parallel to the easy axis (EA) and hard axis (HA). (c) *M*(*H*) loop of a complete TSV which is dominated by the magnetization from the 120-nm-thick LCMO and hence the Py loop (EA) reproduced from (b) is shown for comparison.

### III. RESULTS & DISCUSSION

Figure 2(b) shows $T_C(\theta)$ for a TSV in which the Au layer has not been etched. Comparing P- and AP-states, we see a standard (albeit small) singlet spin-valve effect with $T_C(AP)- T_C(P)$ close to 10 mK. For angles in the 0° < θ < 180° range, $T_C(\theta)$ decreases to a local minima of 5.32 K, close to θ = 60° giving a maximum $T_C$ suppression (defined as $\Delta T_C(\theta) = T_C(AP)-T_C(\theta)$) of -28 mK, which is smaller than the average superconducting transition width. To check that $T_C(\theta)$ cannot be attributed to potential effects arising from field non-uniformity on $T_C$ as the TSV is rotated in-plane during measurements of *R(T)* (e.g. if the sample is not mounted perfectly parallel to the applied field), we investigated $T_C(\theta)$ of the Nb(25nm)/Cu(5nm)/Py(3.5nm)/Au(5nm) control sample with the field applied in-plane and tilted out-of-plane by 10° (see Fig. 2S). A maximum $\Delta T_C(\theta)$ of 10 mK (matching the temperature stability of our system) was observed with no dependence of $T_C$ on θ, meaning that the functional form of $\Delta T_C \theta)$ in Fig. 2(b) is related to the relative magnetizations of Py and LCMO and not field non-uniformity.

The small maximum value of $\Delta T_C(\theta)$ (-28 mK) seen in Fig. 2(b) indicates low interfacial transparency at the Py/Au or Au/LCMO interfaces although we note that *R(T)* does not show anomalous features in the superconducting transition, suggesting a homogeneous interfacial resistance (heterogeneous transparency would result in currents paths changing direction through the superconducting transition so as to preferentially flow in superconducting regions). To improve the Py/Au interface, we Ar-ion etched the Au in situ prior to the sputter-deposition of Nb/Cu/Py and investigated $\Delta T_C(\theta)$ on etching time (the Au etch rate is 0.75 ± 0.04 nm/min). The largest $\Delta T_C(\theta)$ of -140 mK (Fig. 2(c)) was achieved for an etch time of 2 minutes with no observable dependence of $T_C$ on $\theta$ for an etch time of 8 minutes. These data indicate that increasing the etch time has the effect of improving the interface transparency between Py and Au with $\Delta T_C(\theta)$ increasing by 110 mK. Simultaneous, etching had the effect of enhancing the singlet spin-valve effect with $T_C$(AP)- $T_C$(P) increasing from 10 mK (without etching) to 40 mK after 2 minutes of etching (Fig. 3). Over etching the Au, however, risks introducing roughness and ferromagnetic coupling between Py and LCMO and so a decrease in $\Delta T_C(\theta)$ beyond a certain etch time is expected (as seen for an etch time of 8 minutes). We note that, we also investigated using Cu as an alternative to Au at the LCMO interface, but only a singlet spin-valve effect was observed ($T_C$(AP)> $T_C$(P)); see Supplemental Materials for further details.

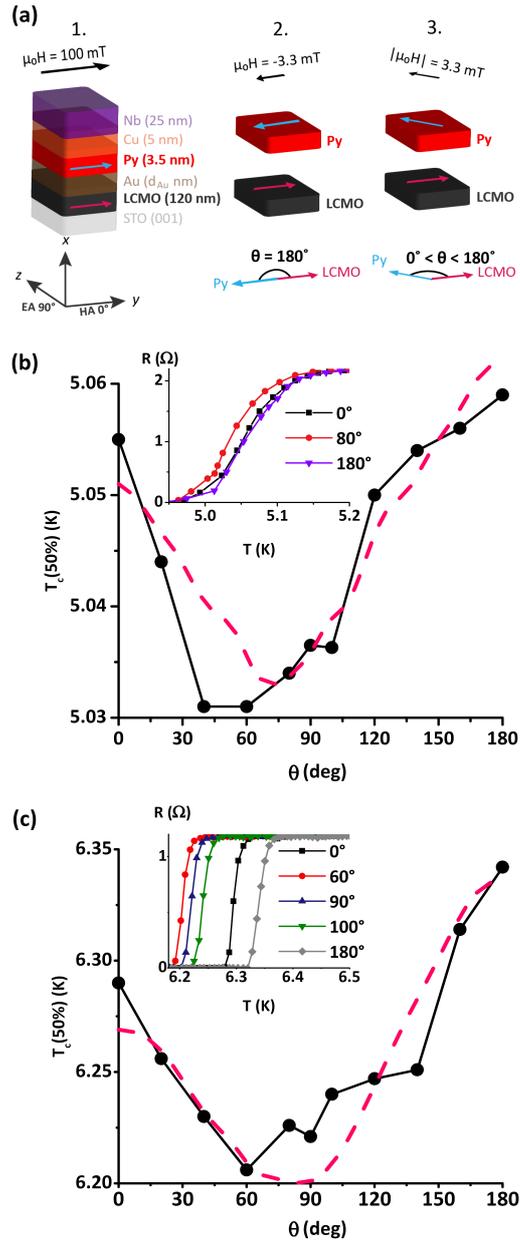

**Fig. 2.** (a) Measurement sequence to measure $T_C$ as a function of $\theta$. The blue and pink arrows show the likely magnetization configuration of Py and LCMO. (b) and (c) show example data of $T_C$(50%) vs $\theta$ for Nb(25nm)/Cu(5nm)/Py(3.5nm)/Au($d_{Au}$)/LCMO(120nm) TSVs without etching of Au ((b); $d_{Au}$=5nm) and following two minutes of etching ((c); $d_{Au}$=3.75nm). The dashed pink lines show the simulated values of $T_C$(50%). The insets shows selected $R(T)$ transitions for various magnetization angles (labelled).

To compare our results to theory, we calculated $\Delta T_C(\theta)$ of the Nb/Cu/Py/Au/LCMO TSVs using a fully microscopic procedure, based on numerical solutions to the self-consistent Bogoliubov-de Gennes (BdG) equations, as extensively discussed in [18, 19, 58]. Each layer is assumed to be infinite in the *y-z* plane (see Fig. 2(a)). The four interfaces between Nb and LCMO will have differing transparencies and to account for spin-independent scattering at these interfaces, we include repulsive delta function potentials $H_i\delta(x-x_i)$ at each interface position $x_i$ (where $i$ = 1-4 refers to the interface number: $i$ = 1 corresponds to the Nb/Cu interface while $i$ = 4 the Cu/LCMO interface). The scattering strength is parameterized in dimensionless units by the quantity $H_{Bi}$, written as $H_{Bi}=mH_i/k_F$, where $k_F$ is the Fermi wavevector, and $m$ is the effective mass. Thus, increasing $H_{Bi}$ decreases the interface transparency[18, 19]. To effectively characterize the TSV and maintain a tractable parameter space, it is necessary to keep the scattering strength combinations as simple as possible. We found good correlation with experiment when setting $H_{B1}$ = $H_{B3}$ = 0.2 for the Au/LCMO and Cu/Py interfaces respectively. For the unetched TSV in (b), we assume a lower transparency at the Py/Au interface with $H_{B2}$ = 1.2, while the Nb/Cu interface is represented with $H_{B4}$ = 0.14. Using these optimised parameters, the model is able to capture the experimental $T_C(\theta)$ behavior seen in Fig. 4 where the local minima in $T_C$ theoretically relates to the transfer of spin-polarized triplet pairs to LCMO (see also Supplementary Material).

It is interesting to note that the experimental and theoretical minimum in $T_C(\theta)$ are shifted from the orthogonal magnetic configuration (θ = 90°). Properly accounting for proximity effects can alter the traditional simple view of the triplet spin valve, whereby the equal spin triplet components undergo a maximum at 90 degrees (leading to a corresponding dip in $T_C$). By including interface scattering, the quasiparticle amplitudes can undergo phase shifts that push the minimum in $T_C$ away from 90°. The same effect also arises in the ballistic regime [18] from the superposition of quasiparticle interactions with the interfaces and outer system walls that causes equal spin triplet pair amplitudes to be largest at relative magnetization angles away from 90°. See also [58] where similar effects are found in the diffusive regime.

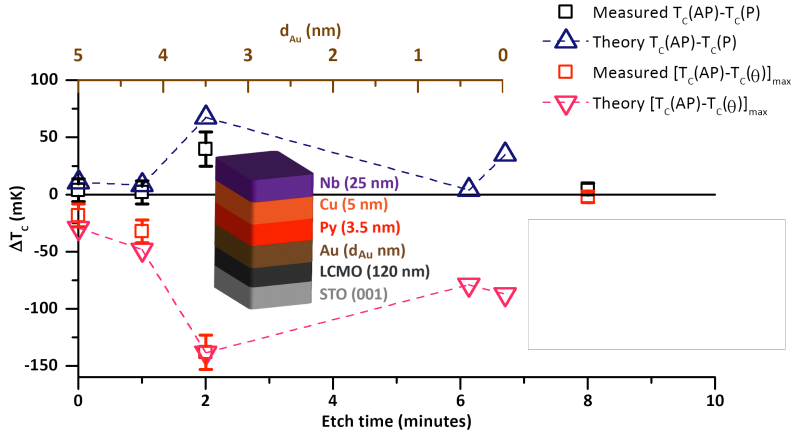

**Fig. 3.** Theory and experimental $\Delta T_C$ vs etch Au etch time and Au layer thickness.

In Fig. 3, we have compared the experimental and calculated dependence of the maximum value of $\Delta T_C(\theta)$ as a function of etching time. To focus on the effect of etching time on the Py/Au interface, we fix all interface scattering parameters, except $H_{B2}$ (relating to the Py/Au interface) which is allowed to vary in such a way that is consistent with the measured etch rate. Namely, we set $H_{B1}$ = $H_{B3}$ = 0.4, $H_{B4}$ = 0.14, and 0.7 ≤ $H_{B2}$ ≤ 1.2. After a certain time, continued etching is assumed to have no further effect on the interface scattering parameter $H_{B2}$. The thickness of the Au, however, decreases (0.75 nm/min) with etching. For each datum point, we self consistently calculate $T_C(\theta)$ and extract $\Delta T_C$ and $T_C$(AP)-$T_C$(P). This results in good agreement with the experimental findings. In particular, the spin valve effect is enhanced for an etching time of 2 minutes whereby an increased singlet-to-triplet pair conversion takes place. Since the normal metal layers tend to host spin-polarised triplet pairs, reducing their thickness can also result in a limited $T_C$ reduction that signifies the emergence of spin polarized triplet pairs.

## IV. SUMMARY

We have demonstrated triplet pair creation through magnetization control in Nb/Cu/Py/Cu/LCMO TSVs using in-plane magnetic field as small as 3.3 mT. Efficient pair conversion and spin-polarized triplet pair transfer to LCMO is achieved for relative magnetization angles between 60° to 90° with a maximum $\Delta T_C(\theta)$ close to -150 mK through band matching optimization at the Au/LCMO interface. Although $\Delta T_C(\theta)$ is smaller than observed for TSVs containing $CrO_2$ which achieved -800 mK [34]) in an out-of-plane magnetic fields of 2 T, our results agree well with a fully microscopic self-consistent model and demonstrate that the fully spin-polarized and chemically stable mixed valance manganites are highly attractive for superconducting spintronics.

### ACKNOWLEDGEMENTS

The work was funded by the Royal Society ('Superconducting Spintronics'), the Leverhulme Trust (IN-2013-033), and the EPSRC through the Programme Grant "Superspin" (EP/N017242/1) and the "International network to explore novel superconductivity at advanced oxide superconductor/magnet interfaces and in nanodevices" Grant (EP/P026311/1) and Doctoral Training Programme (EP/M508007/1). J.W.A.R. and A.D.B. acknowledge support from St John's College, Cambridge. Alidoust is supported by Iran's National Elites Foundation (INEF). K.H. is supported in part by ONR and a grant of HPC resources from the DOD HPCMP.

# Supplemental Materials

**Magnetization-control and transfer of spin-polarized Cooper pairs into a half-metal manganite**

A. Srivastava[1], L. A. B. Olde Olthof[1,2], A. Di Bernardo[1], S. Komori[1], M. Amado[1], C. Palomares-Garcia[1], M. Alidoust[3], K. Halterman[4], M. G. Blamire[1], and J. W. A. Robinson[1]*

[1] Department of Materials Science and Metallurgy, University of Cambridge, 27 Charles Babbage Road, Cambridge CB3 0FS, United Kingdom

[2] Faculty of Science and Technology and MESA+ Institute for Nanotechnology, University of Twente, 7500 AE Enschede, The Netherlands

[3] Department of Physics, K.N. Toosi University of Technology, Tehran 15875-4416, Iran

[4] Michelson Lab, Physics Division, Naval Air Warfare Center, China Lake, California 93555, USA

---

* jjr33@cam.ac.uk

## 1. Structural properties of La$_{2/3}$Ca$_{1/3}$MnO$_3$ (LCMO)

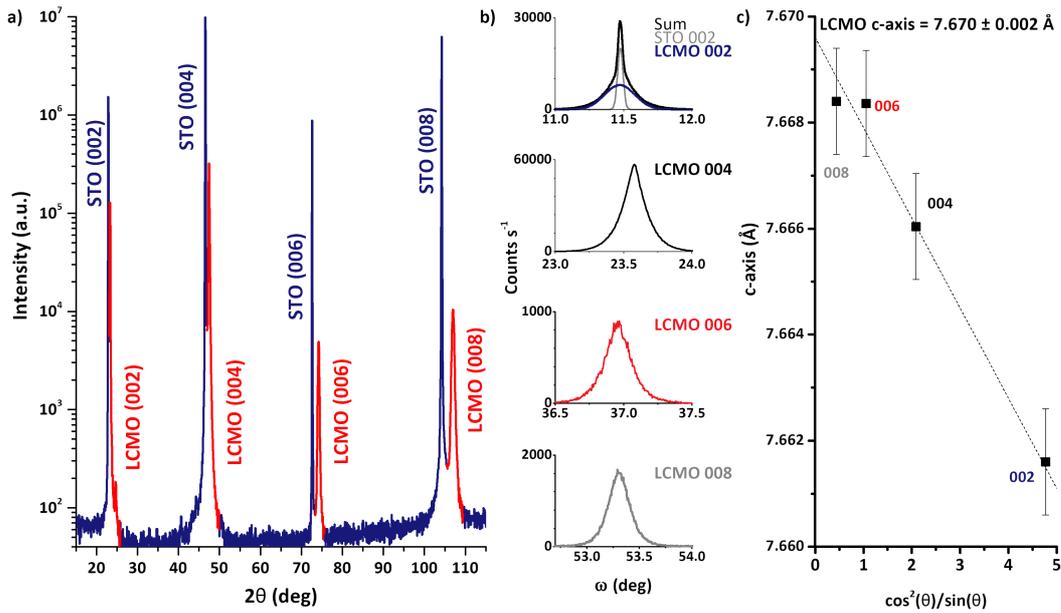

**Fig. 1S.** High angle X-ray diffraction data (left) of 120-nm-thick La$_{2/3}$Ca$_{1/3}$MnO$_3$ (LCMO) on single crystal STO(001), and (right) rocking curves of the LCMO (002), (004), (006) and (008) peaks showing full-width at half maximum values of 0.12°, 0.18°, 0.209° and 0.227° respectively.

## 2. Effect of rotation and tilt angle on the superconducting transition of a control sample

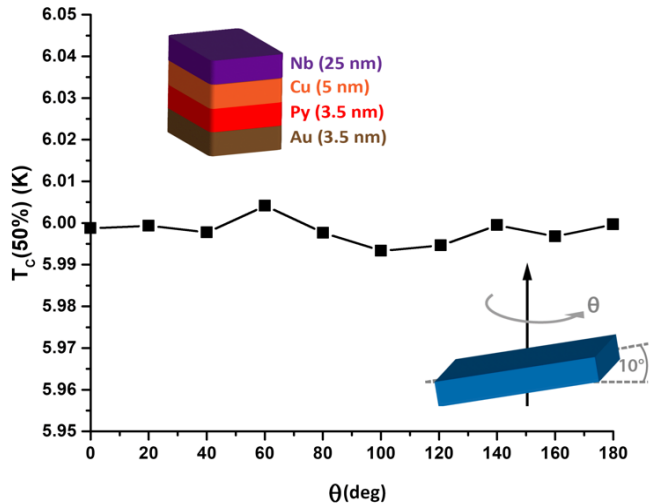

**Fig. 2S.** Superconducting transition $T_C$ of a control sample (sketched in top left inset) as a function of angle and tilted out-of-plane by 10° (as sketch in bottom-right inset).

## 3. Triplet spin-valve with Cu/La$_{2/3}$Ca$_{1/3}$MnO$_3$ (LCMO) interface

We investigated substituting Au (5 nm) with Cu (5 nm). The 120-nm-thick LCMO film was directly transferred to the sputtering chamber and a series of Nb(25nm)/Cu(5nm)/Py(3.5nm)/Cu(5nm) films were deposited following a pre-clean of LCMO by Ar-ion etching using etching times of 0, 2, 3 and 5 minutes. In Fig. 3S we have plotted extracted values of $\Delta T_C(\theta)$ and $T_C$(AP)-$T_C$(P) as a function of etching time. The $\Delta T_C(\theta)$ does not show a dip at intermediate values of θ despite an enhancement of $T_C$(AP)- $T_C$(P) with etching from 12 mK (no etching) to 50 mK (after 5 minutes of etching). These data indicate that, although the Cooper pairs experience the exchange field from LCMO, since $T_C$ is sensitive to P- and AP-alignments of Py and LCMO, triplet pairs are not able to transfer across the Cu/LCMO interface as no suppression of $T_C$ with θ is observed.

The potential blocking of triplet pairs at the Cu/LCMO interface suggests that the Cu/LCMO interface is highly resistive, most likely due to partial oxidation of Cu. Unlike Au, the Gibbs free energy for oxidation of Cu is negative, meaning a thin CuO layer is energetically favourable [59]. Furthermore, CuO will lead to increased oxygen vacancies at the surface of LCMO and a residual $Mn^{2+}$ layer, thus reducing the $Mn^{3+}/Mn^{4+}$ double exchange mechanism responsible for ferromagnetism in LCMO [55]. Since the standard spin singlet effect is observed, we conclude nevertheless that the Cu/LCMO interface is magnetic although the barrier height is too high for triplet pair transfer to LCMO.

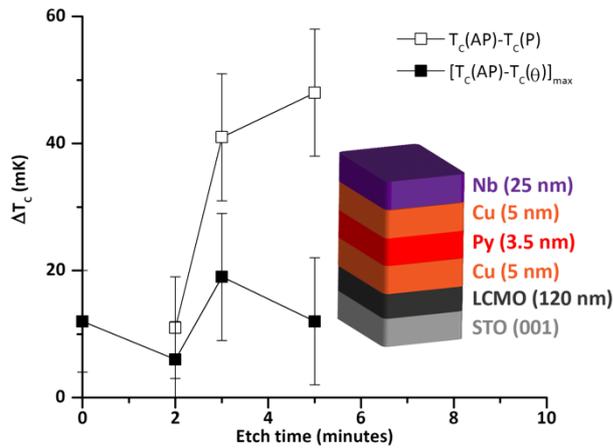

**Fig. 3S.** Experimental $\Delta T_C$ (defined in the legend) as a function of etch time for Nb/Cu/Py/Cu/LCMO triplet spin-valves.

## 4. Isotropic magnetic properties of La$_{2/3}$Ca$_{1/3}$MnO$_3$ (LCMO)

The magnetic properties of a bare 120-nm-thick LCMO thin film was investigated at 10 K as a function of in-plane magnetic field angle. In Fig. 4S we have plotted magnetization versus in-plane applied field $M(H)$ where $M(H)$ is virtually independent of magnetic field orientation with the coercivity ($H_c$) and remanent moment $M_r$ normalized by the saturation magnetization $M_s$ therefore field-angle independent.

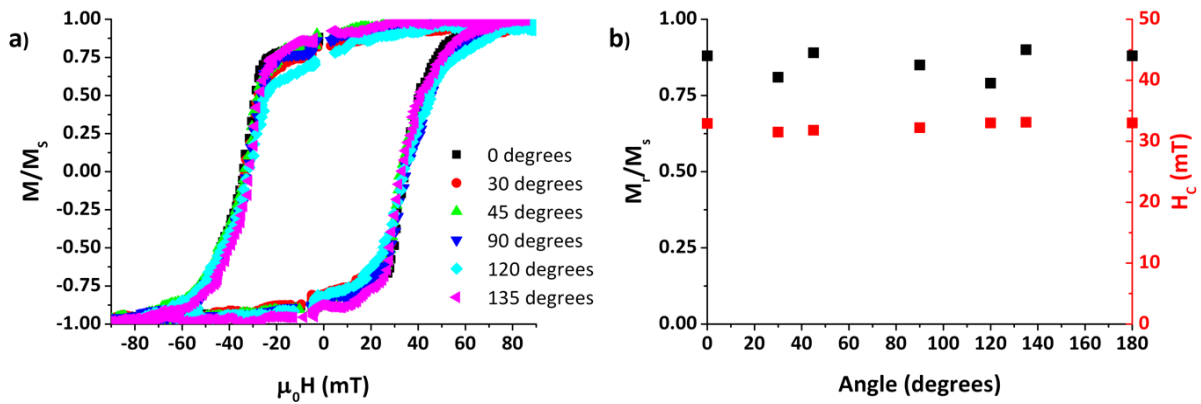

**Fig. 4S.** (a) $M(H)$ loops of bare LCMO (120-nm-thick) at 10 K for different in-plane magnetic field angles (labelled). (b) Remanent moment normalised by the saturation moment ($M_r/M_s$) and coersive field ($H_c$) as a function of in-plane magnetic field angle at 10 K extracted from (a).

## 5. Theoretical model

In general, whenever possible, we set the relevant parameters to their corresponding experimental values. The intrinsic material parameters, e.g., the magnetic exchange field $h$, and coherence length $\xi_0$ are required to be invariant between the samples when modelling the triplet spin-valves. Only the interface scattering parameters, which are expected to vary from sample to sample, are adjusted.

We therefore take the $F$ layer to have a set width of 3.5 nm and variable magnetization direction, while the HMF has a fixed direction of magnetization (along $z$). Due to the existence of one spin band at the Fermi level and strong pair breaking effects, no differences are found in the simulations for HMF widths that exceed 15 nm thick. Thus, to reduce the computation times, we take this minimum thickness of the HM. The $N_1$ layer has a fixed width of 5 nm while the width of $N_2$ can vary due to the effects of etching, but it never exceeds 5 nm. The thin $S$ layer has width 25 nm and an effective correlation length of $\xi_0$ = 17.5 nm. The in-plane magnetizations in the $F$ and HMF layers are modelled by effective Stoner-type exchange fields that vanish within the $N$ and $S$ layers. For the $F$ and HMF layers, we set $h_F$ = 0.05, and $h_{HM}$ = 1 (both in units of $E_F$), respectively. We also set $T_{0c}$ = 8.1 K, corresponding to the critical temperature of a bulk Nb sample. A suitable value of the Fermi wavevector is found to be equal to $k_F$ = 1 Å.

The strength of the triplet spin valve effect, $\Delta T_{C,max}$, is strongly influenced by the scattering parameters, and most importantly $H_{B2}$. Thus, to achieve the largest spin valve effect possible for the given structure, it is desirable to minimize $H_{B2}$ as much as possible. This is seen in the etched case Fig. 2(c), where the $H_{B2}$ interface scattering parameter is expected to be reduced, as the experimental data reveals a much larger $\Delta T_{C,max}$. Hence we set $H_{B2}$ = 0.05, resulting in good agreement between the experimental and theoretical sets. The overall scale differences in Fig. 2(b) and (c) cannot be accounted for by adjusting $H_{B2}$ alone, and so to increase the overall scale of $T_C$ as seen in Fig. 2(c), we allow some variance in the interface transparency at the Nb interface. By decreasing the transparency (increasing $H_{B4}$), proximity effects become diminished, and the leakage of Cooper pairs into the other layers declines. To therefore account for the experimental differences in magnitudes between samples, we set $H_{B4}$ equal to $H_{B4}$ = 0.48. This is similar to allowing the $S$ width to increase so that it is consistent with its experimental uncertainty. Either way, the intrinsically non-superconducting layers have less of an effect on the overall superconductivity in S which becomes effectively more isolated. Regarding the normal metal layers, although $T_{C,max}$ and $\Delta T_{C,max}$ are highly sensitive to the $S$ and $F$ layer widths, we find $\Delta T_{C,max}$ is much less sensitive to changes in the $N_2$ width.

To gain insight into the influence of the HMF layer on $T_C$, we compare the variations of critical temperature against the misalignment angle θ at differing values of the exchange field in $F_2$ layer. In order to be consistent with the experiments, we consider a S/N1/F1/N2/F2 spin valve and set the parameters fixed at those describing the Nb/Cu/Py/Au/LCMO samples in the main text and vary the exchange field of the $F_2$ layer. As seen in Fig. 5S, the critical temperature shows maximum variations when $h_2$ = $E_F$ which is corresponding to HMF [18]. Note that this is consistent with the previous works where triplet spin valves are made of ferromagnets only [32]. The inset panel illustrates the maximum critical temperature change (corresponding $\Delta T_{C,max}$ = $T_{C,max}$ - $T_{C,min}$) as a function of $h_2$. Thus, it is clearly seen that $\Delta T_{C,max}$ is highest when the outer magnetic layer i.e. $F_2$ reaches HMF [18,32]. Theoretically, it is demonstrated that the maximum variation of $T_C$ is directly linked to the spin-polarized Cooper pairs [18]. The HMF highly suppresses the singlet and spin-mixed triplet states and it is maximal when the misalignment angle is almost ~ 90° depending on the quality of interfaces [18,19].

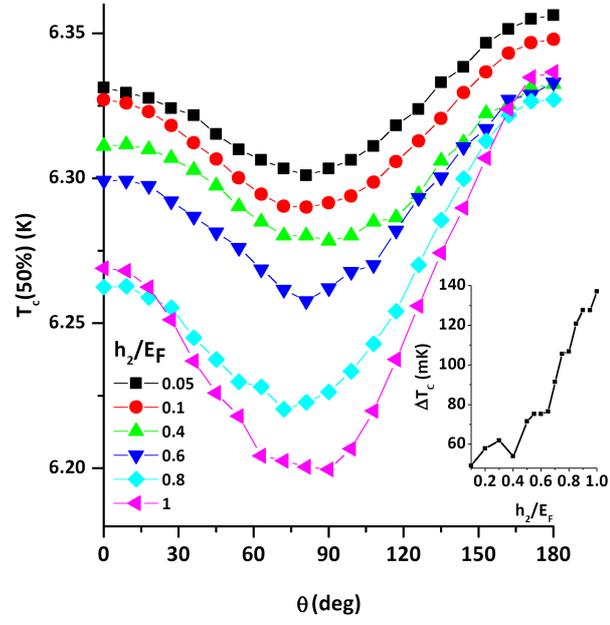

**Fig. 5S.** The critical temperature as a function of magnetization misalignment in the $S/N_1/F_1/N_2/F_2$ spin valve corresponding to the Nb/Cu/Py/Au/LCMO samples. The valve parameters are assumed the same as those found consistent with the experiment and now $h_2$ varies from a weak ferromagnet 0.05 to a HMF 1.0. The inset shows $\Delta T_C$ as a function of $h_2$. We clearly see that the maximum variation in $T_c$ appears when $F_2$ is HMF with $h_2 = E_F$.